\documentclass[]{article}

\usepackage{url}
\usepackage{comment}
\usepackage{todonotes}
\usepackage[utf8]{inputenc} 

\usepackage{graphics}
\usepackage{graphicx}

\title{Using Dual-Network Analyser for extracting communities from Dual Networks}

\author{Pietro Hiram Guzzi, Giuseppe Tradigo, Pierangelo Veltri}                          

\begin{document}
\maketitle

\begin{abstract} 

The representation of data and its relationships using networks is prevalent in many research fields such as computational biology, medical informatics and social networks.  Recently, complex networks models have been introduced to better capture the insights of the modelled scenarios. Among others, dual networks -based models have been introduced, which consist in mapping information as pair of networks containing the same nodes but different edges.   

We focus on the use of a novel approach to visualise and analyse dual networks. The method uses two algorithms for community discovery, and it is provided as a Python-based tool with a graphical user interface. The tool is able to load dual networks and to extract both the densest connected subgraph as well as the common modular communities. The latter is obtained by using an adapted implementation of the Louvain algorithm.

 The proposed algorithm and graphical tool have been tested by using social, biological, and co-authorship networks. Results demonstrate that the proposed approach is efficient and is able to extract meaningful information from dual networks. Finally, as contribution, the proposed graphical user interface can be considered a valuable innovation to the context. 
\end{abstract}


%




\section*{Background}

The use of network-based models is a very popular strategy to analyse data interaction and relations for many domains. E.g., in computational biology, network-based models are used to study relationships among biological macromolecules, and their associations \cite{Cannataro2010,di2015integrated}. In medicine, networks have been used to relate patients \cite{guzzi2020biological,loscalzo2017network} and model possible similarity or conditions. Even social networks data can be modelled using graphs and analysed to extract relevant information regarding connections among users \cite{sapountzi2018social}.

Many approaches of modelling are based on the use of a {\textit single network}, i.e., the use of a single set of nodes and edges to represent data and the subsequent investigation of networks properties such as community-related structures  \cite{cho2013m,cannataro2013data,gu2018homogeneous}. In biology, proteins and their biochemical associations are modelled using the Protein Interaction Network (PINs) formalism. In such a scenario, communities in PINs represent protein complexes, i.e., a set of proteins bound together for performing a specific role \cite{cannataro2010impreco}. 

In order to better capture some aspects real scenarios, the use of a pair of graphs has been introduced. Thus, graph pairs can be used to represent two different views of the same use case, for modelling studying phenomena and evolving scenarios \cite{Wu:2016tx,milano2020hetnetaligner}. In the case of graph pairs, such a model can be indicated as a dual network (DN).
We here focus on the use of \textit{dual networks (DNs)} in which one of the graphs is unweighted and referred to as a physical graph; the other one, called conceptual graph, is edge-weighted. The two graphs may have different (but overlapping) node sets, while in many cases, node sets are the same. 
DNs are used for modelling two different relationships among the nodes, which cannot be modelled using a single graph.
Dual networks find their natural application whenever two kinds of relations among a set of nodes have to be modelled \cite{dondi2020top}. The two networks represent physical and conceptual interactions \cite{guzzi2020extracting,Phillips:2008dm,agapito2018dietos,Tornow:2003kc,Ulitsky:2007iz,cannataro2005preprocessing,antonelli2019integrating}.

Adopting a DN instance to model real data allows us to study interesting network properties through the use of algorithms from graph theory. For instance, a Densest Connected Subgraph (DCS) in a dual network represents a set of nodes being strongly related in the conceptual network and connected in the physical one. As proven in \cite{Wu:2016tx} and \cite{guzzi2020extracting}, DCS may represent related users of a social network who are not mutually connected (suggesting friendship links not modelled yet), or having associated genes/proteins (suggesting possible missing interactions).
Similarly, to model (sub)sets of related gene and proteins, a \textit{common modular graph} could be used since it represents the subgraph having maximum
modularity in the conceptual network and being connected in the physical one. Given a community,  we use the definition of modularity as given in \cite{Lefebvre2010} as the ratio of the density of edges inside  the  community with respect to edges outside the community.

Finding  DCS (and common modular graphs) in a given network, is an NP-hard problem \cite{Wu:2016tx} in its general formulation. 
Techniques can be used to reduce the problem, such as reducing to the set cover problem \cite{Karp:2009ko} or using heuristics. 
Finding the densest graph in a single network can be solved by employing different heuristics; similarly finding a DCS in a dual network, even if it is still a challenging problem, can be solved by using heuristics. For instance, in \cite{Wu:2016tx} authors propose two heuristics based on pruning for solving DCS problem, whereas  in  \cite{guzzi2020extracting} the extraction of DCS from dual networks has been performed using results from the network alignment domain.

We here still focus on the problem of finding DCS, by improving previously obtained results by extending the approach for finding modular communities in DNs. We model the problem as a local network alignment problem, and we propose: (i) a novel algorithm and (ii) a graphical user interface to manage such data.
The proposed approach has two main steps: (i) first, we merge the two input networks into a single alignment graph, (ii) we use an ad-hoc heuristic for extracting densest communities, and the Louvain algorithm \cite{blondel2008fast} for the modular communities. 
The Louvain method is a greedy optimisation method that shows good performances in extracting communities in large graphs by optimising modularity.  Modularity is a value that measures the relative density of edges inside communities with respect to edges outside communities. Briefly the  Louvain method consists in finding small communities then each small community is mapped into one node of a new graph, and the first step is repeated until the method covers all the nodes in the small community. 

Moreover, since there is a lack of tools for analysis, we propose a novel graphical tool, named  \textit{Dual Network Analyser}, able to support the user to extract and visualise DCS and modular communities from an input dual network.
We implemented such an algorithm, and we show the effectiveness of our approach presenting three case studies: (\textit{i}) the first one based on social networks data, (\textit{ii}) the second one based on biological networks and (\textit{iii}) the third one based on a co-authorship network. Experimental results confirm the effectiveness of our approach.

\subsection*{Related Work}

Given a graph mapping real network communitiies, detecting subgraphs having some properties is a key area in graph analysis \cite{lee2010survey,khuller2009finding}. We focus in the literature context of finding the Densest Connected Subgraph (DCS) and modular connected subgraphs in a dual network.
The detection of dense components of a graph found applications in many important fields such as social network analysis \cite{parthasarathy2011community,ma2017detection,hu2005mining}; nevertheless, we claim that a correct definition of \textit{density} is crucial for this problem. Indeed, there exist many definitions leading to the development of different algorithms. One of the first definitions of dense sub-graph is a fully connected sub-graph, i.e. a clique. However the identification of a maximal clique, also referred to as the \textit{maximum clique problem}, is NP-hard \cite{hastad1996clique}, and it is difficult to approximate \cite{bomze1999maximum}.

Wu et al. proposed an algorithm for finding densest connected sub-graph in a dual network \cite{Wu:2016tx}. The algorithm adopts a two step strategy: first it examines the dual network and it prunes the network by eliminating nodes and edges that are not comprised in the optimal solution; then it employs a greedy strategy to find a DCS in the pruned dual network. Our approach is more flexible with respect to this algorithm since we admit more flexibility in the search of DCS, we also find modular communities and, finally, we provide a Graphical User Interface.

The problem of finding a  \textit{densest} sub-graph in a sub-graph has been solved using different heuristics. For instance, Goldberg et al.,  \cite{goldberg1984finding} proposed an algorithm based on maximum-flow approach. Asashiro et al. proposed a greedy algorithm based on the strategy of deleting the node with minimum degree \cite{asahiro2000greedily}. Our method includes also an heuristic by implementing a similar approach but we improve it by extending the method to weighted graphs. There also exist some variants of this problem, i.e. finding the k (overlapping) subgraphs having biggest density \cite{dondi2020top,dondi2021top}. In particular in Dondi et al., \cite{dondi2020top} author proposed a methodology for extracting top-k connected overlapping densest subgraphs in dual networks. Our approach cannot extract overlapping subgraph but it can analyse modular communities and researcher may extend the proposed tool to insert other algorithms.




\section*{Implementation}

We developed a tool called Dual Network Analyser.  The architecture is constituted by the following modules: \begin{itemize}
    \item {\bf Graphical User Interface.} This module is responsible for the interaction with the user. It supports users to define parameters useful to the algorithm executions.
    
    \item {\bf Network Input/Output.} This module is responsible for the reading network from input files, for managing network representation during the execution and for writing output in files. 
    \item \textbf{Graph Alignment.} This module is responsible for the alignment of the physical and conceptual network. 
    \item \textbf{Community Extraction.} This module is responsible for extracting communities from the alignment graph.
\end{itemize}

\subsection*{Algorithm}

The algorithm receives as input: (\textit{i}) two networks (representing a dual network) and a (\textit{ii}) list of nodes to be mapped. Both networks are initially merged together into a single graph, called Weighted Alignment Graph. Each node of the alignment graph represents a pair of nodes of the input network. Edges are added considering the two input networks. Then in order to extract the densest sub-graph of the alignment graph, the Charikar algorithm is used in the case of DCS, while the Louvain algorithm is used in the case of modular communities. Each extracted sub-graph of the alignment graph represents a connected sub-graph of the unweighted networks and a sub-graph of the conceptual network with the given properties (i.e., density or modularity).


\subsection*{Implementation of the proposed modules.}

The tool has been implemented in Python and each module is based on ad hoc realised modules or on the integration of existing libraries as explained in the following:

\begin{itemize}
    \item {\bf Graphical User Interface.} It is based on tkinter python library \cite{grayson2000python} while the  visualisation of graphs  is made possible by wrapping the Netwulf opensource library \cite{aslak2019netwulf}. Netwulf is an interactive visualisation library that is compatible with networkX data-structures. 
    \item {\bf Network Input/Output.}This module is based on the open-source NetworkX library \cite{hagberg2008exploring, dietos} that is package able to create and efficiently manipulate networks.
    \item \textbf{Graph Alignment.} This module is responsible for the alignment of the physical and conceptual network. We wrapped libraries previously developed in \cite{guzzi2020extracting} and available online\footnote{ \url{https://codeocean.com/capsule/7601009/tree}}. 
    \item \textbf{Community Extraction.}  It is based on our implementation of the Charikar algorithm, also available online\footnote{ \url{https://codeocean.com/capsule/7601009/tree}} and on the implementation of the Louvain Algorithm \cite{blondel2008fast} of the cdlib Python Library\footnote{\url{https://cdlib.readthedocs.io/en/latest/}}.
\end{itemize}

\section*{Results}

\subsection*{Using Dual Network Analyser for Analysing Dual Networks}

In order to run an experiment, the user has to launch the software. Then the GUI shows all the parameter that should be inserted to run an experiment. The parameters are: \begin{itemize}
    \item \texttt{Physical Network}: The user has to select the input unweighted network stored as edge list file.
   \item \texttt{Conceptual Network}: The user has to select the input weighted network stored as an edge list file.
   \item \texttt{Similarity File}: The user has to select the file containing the mapping of the nodes among the networks.
    \item \texttt{Delta}: The user has to select the $\delta$ parameter of the algorithm.
\end{itemize}


We measured the time needed for extracting both DCS and modular communities, considering a set of dual input networks with a growing number of both nodes and edges. The total time for the execution $T_{all}$ is given by the sum of loading the network $T{load}$, calculating the weighted alignment graph $T_{align}$ and extracting communities, $T_{dcs}$ and $T_{com}$ respectively when DCS or Louvain algorithm is used. We considered the networks whose parameter are summarised in Table \ref{tab:partime}.


\subsection*{Extracting Modular Communities from Dual Networks.}

We built a dual-network considering both physical and conceptual interactions among proteins. We selected the STRING database \cite{stringdatabase} containing data related to functional associations among proteins and the I2D \cite{kotlyar2016integrated} database containing data related to physical interactions.

We built two networks:
\begin{itemize}
\item a conceptual network, which represents the strength of associations among proteins extracted from STRING database;
\item a physical network, which stores the binary interactions among proteins extracted from I2D database.
\end{itemize}

We obtained two networks having 19.354 nodes and 5.879.727 edges. We used our tools to extract communities (using $\delta=4$ - these parameters showed best performances)) , and it resulted in 25 top modular communities.   We performed a biological interpretation of the results by using a functional enrichment algorithm provided by the DAVID software \cite{da2007david}.



\section*{Discussion}

\subsection*{Case Study}


In this section, we show how the algorithm is able to recover modular communities as proof of principle. We already showed the benchmark of the densest common subgraph discovery in \cite{guzzi2020extracting}.

We demonstrate that our findings have superior quality over other classical  approaches. The quality of the results is evaluated in various ways: we first show the ability of our approach to recover known modular  by means of the measures of precision and recall, then we show that our solutions are better than other methods.

\subsubsection*{Proof of Concept for Louvain Algorithm. }

We build 100 test dual networks each one containing  Com ($Com_{kn,i}, i=1..200$). Each physical network has 500 nodes and 3000 edges and the conceptual network has 500 nodes and 4000 edges. 

The quality of a  result was evaluated by comparing each extracted Com ($Com_{ex,j}$) with each known $Com_{kn,i}$. The Com sensitivity ($Sn_{Com}$) represents the coverage of a known community by its best-matching extracted community (the maximal fraction of nodes in the community found in a common extracted community). Reciprocally, the Com-wise Positive Predictive Value ($PPV_{Com}$) measures how well a given extracted community predicts its best-matching known community.

To estimate the overall correspondence between a result (a set of extracted modular communities) and the collection of known modular communities, we computed the weighted means of all PPV values (averaged over all extracted communities) and $Sn_{Com}$ values (averaged over all known communities). The resulting statistics, clustering-wise PPV and clustering-wise Sn, provide information about the quality.  To integrate the two measures, we computed a geometrical accuracy ($Acc_{Com}$), defined as the geometrical mean of the averaged Sn and PPV values.

Since classical Louvain algorithm does not run in the dual network, we applied it over conceptual networks; then, we derived the induced sub-graph into the physical network. Finally,  we reduced the cluster on the conceptual network to find a connected sub-graph into the physical one.

We used the  Louvain algorithm on the conceptual network, and we compared with respect to the Louvain algorithm in our framework. Table \ref{tab:results} summarises the performance of the algorithm measured by using the average value evaluated on the runs over each of the 100 networks, respectively, for  PPV, SSN and ACC.

As evidence, we averaged over 100 networks outperforms the classical Louvain Algorithm. 


\section*{Conclusions}


Dual networks are composed of a pair of graphs: an unweighted one (physical network) and an edge-weighted one (conceptual network). We presented a framework composed of a software platform and a comprehensive set of heuristics for the analysis of dual networks. The software is able to obtain both the densest connected sub-graph (DCS) ( having the largest density in the conceptual network and being also connected in the physical network) as well as the common subgraph with the highest modularity. We formalised the problem,  we proposed a possible solution and presented a software with set of experiments, which demonstrate the effectiveness of our approach.

\section*{Acknowledgements}
This work has been partially funded by PON-VQA project.

\section*{Availability and requirements}

\begin{itemize}
    \item Project name: e.g. DN-Analyzer
     \item Project home page: e.g. https://github.com/hguzzi/DNANALYZER
    \item  Operating system(s): e.g. Platform independent
    \item  Programming language: Python 3
    \item  Other requirements: Python 3.7 or higher, Networkx, 
   \item   License: e.g. GNU GPL
   \item   Any restrictions to use by non-academics: Non Commercial Use Only, CC-BY

\end{itemize}





\section*{Competing interests}
  The authors declare that they have no competing interests.

\section*{Author's contributions}

All the authors participated to the design of the proposed software.
All the authors participated to the writing of the manuscript.
    All the authors read and approved the manuscript.

\section*{Acknowledgements}
  Authors thank Eng. Emanuel Salerno for his work on developing software modules.

\bibliographystyle{plain} 
\bibliography{a.bib}      

\begin{thebibliography}{10}

\bibitem{agapito2018dietos}
Giuseppe Agapito, Mariadelina Simeoni, Barbara Calabrese, Ilaria Car{\'e},
  Theodora Lamprinoudi, Pietro~H Guzzi, Arturo Pujia, Giorgio Fuiano, and Mario
  Cannataro.
\newblock Dietos: A dietary recommender system for chronic diseases monitoring
  and management.
\newblock {\em Computer methods and programs in biomedicine}, 153:93--104,
  2018.

\bibitem{dietos}
Giuseppe Agapito, Mariadelina Simeoni, Barbara Calabrese, Ilaria Car{\'e},
  Theodora Lamprinoudi, Pietro~H Guzzi, Arturo Pujia, Giorgio Fuiano, and Mario
  Cannataro.
\newblock Dietos: A dietary recommender system for chronic diseases monitoring
  and management.
\newblock {\em Computer methods and programs in biomedicine}, 153:93--104,
  2018.

\bibitem{antonelli2019integrating}
Laura Antonelli, Mario~Rosario Guarracino, Lucia Maddalena, and Mara
  Sangiovanni.
\newblock Integrating imaging and omics data: A review.
\newblock {\em Biomedical Signal Processing and Control}, 52:264--280, 2019.

\bibitem{asahiro2000greedily}
Yuichi Asahiro, Kazuo Iwama, Hisao Tamaki, and Takeshi Tokuyama.
\newblock Greedily finding a dense subgraph.
\newblock {\em Journal of Algorithms}, 34(2):203--221, 2000.

\bibitem{aslak2019netwulf}
Ulf Aslak and Benjamin~F Maier.
\newblock Netwulf: Interactive visualization of networks in python.
\newblock {\em Journal of Open Source Software}, 4(42):1425, 2019.

\bibitem{blondel2008fast}
Vincent~D Blondel, Jean-Loup Guillaume, Renaud Lambiotte, and Etienne Lefebvre.
\newblock Fast unfolding of communities in large networks.
\newblock {\em Journal of statistical mechanics: theory and experiment},
  2008(10):P10008, 2008.

\bibitem{bomze1999maximum}
Immanuel~M Bomze, Marco Budinich, Panos~M Pardalos, and Marcello Pelillo.
\newblock The maximum clique problem.
\newblock In {\em Handbook of combinatorial optimization}, pages 1--74.
  Springer, 1999.

\bibitem{cannataro2010impreco}
Mario Cannataro, Pietro~H Guzzi, and Pierangelo Veltri.
\newblock Impreco: Distributed prediction of protein complexes.
\newblock {\em Future Generation Computer Systems}, 26(3):434--440, 2010.

\bibitem{Cannataro2010}
Mario Cannataro, Pietro~H. Guzzi, and Pierangelo Veltri.
\newblock {Protein-to-protein interactions}.
\newblock {\em ACM Computing Surveys}, 43(1):1--36, November 2010.

\bibitem{cannataro2005preprocessing}
Mario Cannataro, Pietro~Hiram Guzzi, Tommaso Mazza, Giuseppe Tradigo, and
  Pierangelo Veltri.
\newblock Preprocessing of mass spectrometry proteomics data on the grid.
\newblock In {\em 18th IEEE Symposium on Computer-Based Medical Systems
  (CBMS'05)}, pages 549--554. IEEE, 2005.

\bibitem{cannataro2013data}
Mario Cannataro, Pietro~Hiram Guzzi, and Alessia Sarica.
\newblock Data mining and life sciences applications on the grid.
\newblock {\em Wiley Interdisciplinary Reviews: Data Mining and Knowledge
  Discovery}, 3(3):216--238, 2013.

\bibitem{cho2013m}
Young-Rae Cho, Marco Mina, Yanxin Lu, Nayoung Kwon, and Pietro~H Guzzi.
\newblock M-finder: Uncovering functionally associated proteins from
  interactome data integrated with go annotations.
\newblock {\em Proteome science}, 11(1):S3, 2013.

\bibitem{da2007david}
Brad T~Sherman Da~Wei~Huang, Qina Tan, Jack~R Collins, W~Gregory Alvord, Jean
  Roayaei, Robert Stephens, Michael~W Baseler, H~Clifford Lane, and Richard~A
  Lempicki.
\newblock The david gene functional classification tool: a novel biological
  module-centric algorithm to functionally analyze large gene lists.
\newblock {\em Genome biology}, 8(9):R183, 2007.

\bibitem{di2015integrated}
Maria~Teresa Di~Martino, Pietro~Hiram Guzzi, Daniele Caracciolo, Luca Agnelli,
  Antonino Neri, Brian~A Walker, Gareth~J Morgan, Mario Cannataro,
  Pierfrancesco Tassone, and Pierosandro Tagliaferri.
\newblock Integrated analysis of micrornas, transcription factors and target
  genes expression discloses a specific molecular architecture of hyperdiploid
  multiple myeloma.
\newblock {\em Oncotarget}, 6(22):19132, 2015.

\bibitem{dondi2020top}
Riccardo Dondi, Pietro~Hiram Guzzi, and Mohammad~Mehdi Hosseinzadeh.
\newblock Top-k connected overlapping densest subgraphs in dual networks.
\newblock In {\em International Conference on Complex Networks and Their
  Applications}, pages 585--596. Springer, 2020.

\bibitem{dondi2021top}
Riccardo Dondi, Mohammad~Mehdi Hosseinzadeh, Giancarlo Mauri, and Italo Zoppis.
\newblock Top-k overlapping densest subgraphs: approximation algorithms and
  computational complexity.
\newblock {\em Journal of Combinatorial Optimization}, 41(1):80--104, 2021.

\bibitem{goldberg1984finding}
AV~Goldberg.
\newblock Finding a maximum density subgraph. technical report.
\newblock {\em Uni. California, Berkeley}, 1984.

\bibitem{grayson2000python}
John~E Grayson.
\newblock {\em Python and Tkinter programming}.
\newblock Manning Publications Co. Greenwich, 2000.

\bibitem{gu2018homogeneous}
Shawn Gu, John Johnson, Fazle~E Faisal, and Tijana Milenkovi{\'c}.
\newblock From homogeneous to heterogeneous network alignment via colored
  graphlets.
\newblock {\em Scientific reports}, 8(1):1--16, 2018.

\bibitem{guzzi2020biological}
Pietro~Hiram Guzzi and Swarup Roy.
\newblock {\em Biological Network Analysis: Trends, Approaches, Graph Theory,
  and Algorithms}.
\newblock Elsevier, 2020.

\bibitem{guzzi2020extracting}
Pietro~Hiram Guzzi, Emanuel Salerno, Giuseppe Tradigo, and Pierangelo Veltri.
\newblock Extracting dense and connected communities in dual networks: An
  alignment based algorithm.
\newblock {\em IEEE Access}, 8:162279--162289, 2020.

\bibitem{hagberg2008exploring}
Aric Hagberg, Pieter Swart, and Daniel S~Chult.
\newblock Exploring network structure, dynamics, and function using networkx.
\newblock Technical report, Los Alamos National Lab.(LANL), Los Alamos, NM
  (United States), 2008.

\bibitem{hastad1996clique}
Johan Hastad.
\newblock Clique is hard to approximate within n/sup 1-/spl epsiv.
\newblock In {\em Proceedings of 37th Conference on Foundations of Computer
  Science}, pages 627--636. IEEE, 1996.

\bibitem{hu2005mining}
Haiyan Hu, Xifeng Yan, Yu~Huang, Jiawei Han, and Xianghong~Jasmine Zhou.
\newblock Mining coherent dense subgraphs across massive biological network for
  functional discovery.
\newblock {\em Bioinformatics}, 1(1):1--9, 2005.

\bibitem{Karp:2009ko}
Richard~M Karp.
\newblock {Reducibility Among Combinatorial Problems}.
\newblock In {\em 50 Years of Integer Programming 1958-2008}, pages 219--241.
  Springer Berlin Heidelberg, Berlin, Heidelberg, November 2009.

\bibitem{khuller2009finding}
Samir Khuller and Barna Saha.
\newblock On finding dense subgraphs.
\newblock In {\em International Colloquium on Automata, Languages, and
  Programming}, pages 597--608. Springer, 2009.

\bibitem{kotlyar2016integrated}
Max Kotlyar, Chiara Pastrello, Nicholas Sheahan, and Igor Jurisica.
\newblock Integrated interactions database: tissue-specific view of the human
  and model organism interactomes.
\newblock {\em Nucleic acids research}, 44(D1):D536--D541, 2016.

\bibitem{lee2010survey}
Victor~E Lee, Ning Ruan, Ruoming Jin, and Charu Aggarwal.
\newblock A survey of algorithms for dense subgraph discovery.
\newblock In {\em Managing and Mining Graph Data}, pages 303--336. Springer,
  2010.

\bibitem{Lefebvre2010}
Celine Lefebvre, Presha Rajbhandari, Mariano~J Alvarez, Pradeep Bandaru,
  Wei~Keat Lim, Mai Sato, Kai Wang, Pavel Sumazin, Manjunath Kustagi, Brygida~C
  Bisikirska, Katia Basso, Pedro Beltrao, Nevan Krogan, Jean Gautier, Riccardo
  Dalla-Favera, and Andrea Califano.
\newblock {A human B-cell interactome identifies MYB and FOXM1 as master
  regulators of proliferation in germinal centers.}
\newblock {\em Molecular systems biology}, 6(377):377, June 2010.

\bibitem{loscalzo2017network}
Joseph Loscalzo.
\newblock {\em Network medicine}.
\newblock Harvard University Press, 2017.

\bibitem{ma2017detection}
Xiuli Ma, Guangyu Zhou, Jingbo Shang, Jingjing Wang, Jian Peng, and Jiawei Han.
\newblock Detection of complexes in biological networks through diversified
  dense subgraph mining.
\newblock {\em Journal of Computational Biology}, 24(9):923--941, 2017.

\bibitem{milano2020hetnetaligner}
Marianna Milano, Tijana Milenkovi{\'c}, Mario Cannataro, and Pietro~Hiram
  Guzzi.
\newblock L-hetnetaligner: A novel algorithm for local alignment of
  heterogeneous biological networks.
\newblock {\em Scientific reports}, 10(1):1--20, 2020.

\bibitem{parthasarathy2011community}
Srinivasan Parthasarathy, Yiye Ruan, and Venu Satuluri.
\newblock Community discovery in social networks: Applications, methods and
  emerging trends.
\newblock In {\em Social network data analytics}, pages 79--113. Springer,
  2011.

\bibitem{Phillips:2008dm}
Patrick~C Phillips.
\newblock {Epistasis {\textemdash} the essential role of gene interactions in
  the structure and evolution of genetic systems}.
\newblock {\em Nature reviews. Genetics}, 9(11):855--867, November 2008.

\bibitem{sapountzi2018social}
Androniki Sapountzi and Kostas~E Psannis.
\newblock Social networking data analysis tools \& challenges.
\newblock {\em Future Generation Computer Systems}, 86:893--913, 2018.

\bibitem{stringdatabase}
Damian Szklarczyk, John~H Morris, Helen Cook, Michael Kuhn, Stefan Wyder, Milan
  Simonovic, Alberto Santos, Nadezhda~T Doncheva, Alexander Roth, Peer Bork,
  et~al.
\newblock The string database in 2017: quality-controlled protein--protein
  association networks, made broadly accessible.
\newblock {\em Nucleic acids research}, page gkw937, 2016.

\bibitem{Tornow:2003kc}
Sabine Tornow and HW~Mewes.
\newblock Functional modules by relating protein interaction networks and gene
  expression.
\newblock {\em Nucleic Acids Research}, 31(21):6283--6289, 2003.

\bibitem{Ulitsky:2007iz}
Igor Ulitsky and Ron Shamir.
\newblock {Pathway redundancy and protein essentiality revealed in the
  Saccharomyces cerevisiae interaction networks}.
\newblock {\em Molecular systems biology}, 3(1):104, April 2007.

\bibitem{Wu:2016tx}
Yubao Wu, Xiaofeng Zhu, Li~Li, Wei Fan, Ruoming Jin, and Xiang Zhang.
\newblock {Mining Dual Networks - Models, Algorithms, and Applications.}
\newblock {\em TKDD}, 2016.

\end{thebibliography}





\section*{Tables}

\begin{table}[h!]
\caption{Characteristics of the Dual Network used for Test.}
\label{tab:partime}
      \begin{tabular}{|c|c|c|c|c|}\hline
      No& \multicolumn{2}{|c|}{Physical Network}& \multicolumn{2}{|c|}{Conceptual Network} \\ \hline
 Experiment & Nodes & Edges & Nodes & Edges \\ \hline  
1&500 & 1000 & 500 & 1500 \\ \hline
2&1000 & 1500 & 1000 & 2000 \\ \hline
3&1500 & 2000 & 1500 & 2500 \\ \hline
4&2000 & 2500 & 2000 & 3000 \\ \hline
5&2500 & 3000 & 2500 & 3500 \\ \hline
      \end{tabular}
\end{table}

\begin{table}[h!]
\caption{Execution Time}
\label{tab:partime}
      \begin{tabular}{|c|c|c|c|c|}\hline
     Experiment & $T_{load}$ ms & $T_{align}$ ms & $T_{dcs}$ ms & $T_{louv}$ ms  \\ \hline
     1 & 10 & 100 & 150 & 200 \\ \hline 
       2 & 15 & 200 & 250 & 300 \\ \hline 
         3 & 21 & 300 & 350 & 400 \\ \hline 
           4 & 23 & 500 & 450 & 500 \\ \hline 
             5 & 35 & 1900 & 650 & 700 \\ \hline 
      \end{tabular}
\end{table}

\begin{table}[ht]				
    \centering				
    \begin{tabular}{|l|c|c|c|}\hline				
				
Algorithm &	PPV	& SSN	&ACC \\ \hline 	
DN Aligner &	0.70 &	0.78 &	0.73 	\\ \hline
LOUVAIN	& 0.68   &	0.68 	& 0.68 \\	\hline 
    \end{tabular}				
    \caption{Performances on synthetic networks: {\bf average values are reported with their standard deviation}}				
    \label{tab:results}				
\end{table}




\end{document}